\documentclass[pra,twocolumn,tightenlines,showpacs,nofootinbib]{revtex4-1}
\usepackage{bm,dcolumn,amsmath,graphicx,amsfonts,amssymb}
\usepackage{epsfig}
\begin{document} 
\title{Parity nonconservation in Fr-like actinide and Cs-like rare-earth  ions} 
\author{B. M. Roberts}\email[]{b.roberts@unsw.edu.au}
\author{V. A. Dzuba}
\author{V. V. Flambaum}
\affiliation{School of Physics, University of New South Wales, Sydney, NSW 2052, Australia}
\date{ \today }

\begin{abstract}
Parity nonconservation amplitudes are calculated  for the
7$s$-6$d_{3/2}$ transitions of the francium isoelectronic sequence
(Fr, Ra$^+$, Ac$^{2+}$, Th$^{3+}$, Pa$^{4+}$, U$^{5+}$ and Np$^{6+}$)
and for the 6$s$-5$d_{3/2}$ transitions of the cesium isoelectronic
sequence (Cs, Ba$^{+}$, La$^{2+}$, Ce$^{3+}$ and Pr$^{4+}$).  
We show in particular that isotopes of La$^{2+}$, Ac$^{2+}$ and
Th$^{3+}$ ions have strong potential in the search for new physics
beyond the standard model -- 
the PNC amplitudes are large, the calculations are accurate and the
nuclei are practically stable.  
In addition, ${}^{232}$Th$^{3+}$ ions have recently been trapped and
cooled [C. J. Campbell {\em et al.}, Phys. Rev. Lett. {\bf 102},
233004 (2009)].  
We also extend previous works by calculating the $s$-$s$ PNC
transitions in Ra$^{+}$ and Ba$^{+}$, and provide new calculations of
several energy levels, and electric dipole and quadrupole transition
amplitudes for the Fr-like actinide ions. 
\end{abstract}
\pacs{11.30.Er, 31.15.A-, 31.30.jg}
\maketitle


\section{Introduction}

The parity nonconservation (PNC) amplitude of the $6s$-$7s$ transition in cesium is presently
the most precise low-energy test of the electroweak theory.  
This precision is a result of highly accurate measurements~\cite{meas}
and the almost equally accurate  atomic calculations used for their
interpretation~\cite{CsPNC,Porsev,CsPNCour,CsPNCold}.  
This interpretation shows that the value of the weak nuclear charge
for $^{133}$Cs, coming from the PNC measurements, differs from the
prediction of the standard model by 1.6$\sigma$~\cite{CsPNC,CsQw}.  
Although this cannot be regarded as disagreement, it indicates that
further improvements to the accuracy of the measurements and
interpretation may lead to new important results. 

Additionally, the need for new measurements is motivated by the requirement to check the existing results. 
It is very important that an independent test of the existing experimental results~\cite{meas} is performed. These are very important results and must be checked even if the accuracy is not improved. 
Moreover, there is a good chance for both theoretical and experimental improvement since the PNC amplitudes  of these systems are up to 50 times larger than in Cs. 

It is natural to expect a higher accuracy in measurements of systems
where the PNC effect is larger. On the other hand, for high accuracy
of the calculations it is important to have systems with a simple
electron structure. The $s$-$d$ transitions in the Fr-like ions
considered in this paper seem to be very promising in this regard.  
The PNC amplitude is larger for atoms or ions with higher nuclear
charge $Z$~\cite{Khr91}. It is also larger for the $s$-$d$
transitions~\cite{Ba+} than for the $s$-$s$ transitions like the one
used in cesium~\cite{CsPNC}. The accuracy of these calculations is
also expected to be high since the ions have simple electron structure
with one valence electron above closed shells similar to that of cesium.  

There are several additional factors which promise potentially better
theoretical accuracy for these ions than for Cs:  
\begin{itemize}
\item 
The main source of theoretical uncertainty is electron correlations. 
The relative value of the correlation correction is smaller for ions
than for neutral atoms.  
\item 
There are no strong cancellations between different correlation
corrections for $s$-$d$ PNC amplitudes~\cite{Ba+} in contrast to very
strong cancellation for the $6s$-$7s$ PNC amplitude in Cs.
\item 
The $s$-$d$ PNC amplitude is strongly dominated by the term with the
$d$-$p$ electric dipole transition amplitude and $s$-$p$ weak matrix
element. This term can be checked and/or corrected if an accurate
experimental value for the $d$-$p$ amplitude is known.  
A similar approach in Cs works with limited accuracy due to strong
cancellation between $6s$-$np$ and $7s$-$np$ contributions.    
\end{itemize}

PNC measurements have been considered for the Ba$^+$ ion~\cite{BaII}
and are in progress for the Ra$^+$ ion~\cite{KVI}. The FrPNC
collaboration has begun the construction of a laser cooling and
trapping apparatus with the purpose of measuring atomic parity
nonconservation in microwave and optical transitions of
francium~\cite{FrPNC}. With a PNC amplitude in the $7s$-$8s$ optical
transition expected to be around 15 times larger than that of cesium,
and its relatively simple electronic configuration, francium is a very
good candidate atom for precision measurements and calculations of
PNC~\cite{FrDzuba,Ba+,FrSaf}.   

With the aim of motivating experiment in this important area, we
present calculations of $s$-$s$ and $s$-$d$ PNC amplitudes for several
Cs- and Fr-like ions. Simple estimates show that the size of the PNC
effect should scale as $E_{PNC}\sim Z^3 R(Z\alpha)\,/\,Z_a$, where $R$
is a relativistic factor, $Z$ is nuclear charge and $Z_a$ is the
effective charge defined as $E_n= -Z^2 _a/2n^2$ atomic units
~\cite{OngPRA}. Here, $E_n$ is the energy of valence electron, $n$ the
principal quantum number ($n=7$ for Fr-like ions). Therefore PNC
effects in these ions are only slightly smaller than in neutral
atoms. 

Of particular interest are the optical $s$-$d$ PNC transitions of
${}^{232}$Th$^{3+}$ and ${}^{139}$La$^{2+}$, and the IR transition in
${}^{227}$Ac$^{2+}$. ${}^{232}$Th has a half-life of
$1.5\times10^{10}$ years and ${}^{227}$Ac of $21.8$ years, much more
stable than Fr with its most stable isotope (${}^{223}$Fr) having a
half-life of just 22 minutes. ${}^{139}$La$^{2+}$ is stable. 
Importantly, the ${}^{232}$Th$^{3+}$ ion has been trapped and cooled
by Campbell {\em et al.}~\cite{ThLaserCool}. This was the first
reported laser cooling of a multiply charged ion.

The experiment needed to measure the $7s$-$6d_{3/2}$  PNC amplitude in
Th$^{3+}$ is somewhat different to the conventional PNC experiments.  
Neither of the states of interest are the ground state and the PNC
amplitude must be reached by first populating the metastable $7s$
state. This is explored in more detail in the later
sections. Ac$^{2+}$ maintains a $7s$ ground state, and has a very long
lived $6d_{3/2}$ state, which is highly beneficial for PNC
measurements~\cite{BaII}.

${}^{227}$Ac and the odd-nucleon isotope ${}^{229}$Th (with a
half-life of $7340$ years) will also be of interest for measurements
of nuclear-spin-dependent PNC in optical or hyperfine transitions,
including the extraction of the nuclear anapole moment and the
strength of the PNC nuclear forces (see, e.g. \cite{GFrev04}).

\section{Calculations}

\begin{table*}[ht!]%
    \centering%
    \caption{Comparison of calculated energy levels (BO) and experimental values (Ref.~\cite{NIST}) for Cs, Ba${}^{+}$, Fr and Ra${}^{+}$. (cm$^{-1}$)}%
\begin{ruledtabular}%
\begin{tabular}{lrr|lrr|lrr|lrr}
  \multicolumn{3}{c|}{Cs} &       \multicolumn{3}{c|}{Ba${}^{+}$} &       \multicolumn{3}{c|}{Fr} &       \multicolumn{3}{c}{Ra${}^{+}$}   \\
      Level & \multicolumn{1}{c}{BO}  &  \multicolumn{1}{c|}{Exp.}  &       & \multicolumn{1}{c}{BO}  &  \multicolumn{1}{c|}{Exp.}  &       & \multicolumn{1}{c}{BO}  &  \multicolumn{1}{c|}{Exp.}  &      & \multicolumn{1}{c}{BO}  &  \multicolumn{1}{c}{Exp.}  \\
  \hline 
  6s$_{1/2}$  & 0     & 0     & 6s$_{1/2}$ & 0     & 0     &  7s$_{1/2}$  & 0     & 0     &  7s$_{1/2}$  & 0     & 0 \\
  6p$_{1/2}$  & 11168 & 11178 & 5d$_{3/2}$ & 4280  & 4874  &  7p$_{1/2}$  & 12218 & 12237 &  6d$_{3/2}$  & 11741 & 12084 \\
  6p$_{3/2}$  & 11736 & 11732 & 5d$_{5/2}$ & 5128  & 5675  &  7p$_{3/2}$  & 13954 & 13924 &  6d$_{5/2}$  & 13471 & 13743 \\
  5d$_{3/2}$  & 14310 & 14499 & 6p$_{1/2}$ & 20234 & 20262 &  6d$_{3/2}$  & 16200 & 16230 &  7p$_{1/2}$  & 21291 & 21351 \\
  5d$_{5/2}$  & 14426 & 14597 & 6p$_{3/2}$ & 21960 & 21952 &  6d$_{5/2}$  & 16412 & 16430 &  7p$_{3/2}$  & 26259 & 26209 \\
  7s$_{1/2}$  & 18631 & 18536 & 7s$_{1/2}$ & 42647 & 42355 &  8s$_{1/2}$  & 19862 & 19740 &  8s$_{1/2}$  & 43757 & 43405 \\
  7p$_{1/2}$  & 21818 & 21765 & 6d$_{3/2}$ & 46234 & 45949 &  8p$_{1/2}$  & 23190 & 23113 &  7d$_{3/2}$  & 49082 & 48744 \\
  7p$_{3/2}$  & 22000 & 21946 & 6d$_{5/2}$ & 46438 & 46155 &  8p$_{3/2}$  & 23737 & 23658 &  5f$_{5/2}$  & 49254 & 48988 \\
  6d$_{3/2}$  & 22611 & 22589 & 4f$_{5/2}$ & 47829 & 48259 &  7d$_{3/2}$  & 24311 & 24244 &  7d$_{5/2}$  & 49485 & 49240 \\
  6d$_{5/2}$  & 22656 & 22632 & 4f$_{7/2}$ & 48219 & 48483 &  7d$_{5/2}$  & 24402 & 24333 &  5f$_{7/2}$  & 49569 & 49272 \\
  8s$_{1/2}$  & 24391 & 24317 & 7p$_{1/2}$ & 49595 & 49390 &  9s$_{1/2}$  & 25773 & 25671 &  8p$_{1/2}$  & 50864 & 50606 \\
  4f$_{7/2}$  & 24528 & 24472 & 7p$_{3/2}$ & 50213 & 50011 &  5f$_{7/2}$  & 25970 &  ---  &  8p$_{3/2}$  & 52635 & 52392 \\
  4f$_{5/2}$  & 24528 & 24472 & 8s$_{1/2}$ & 58258 & 58025 &  5f$_{5/2}$  & 25971 &  ---  &  9s$_{1/2}$  & 59448 & 59165 \\
  Lim.:\tablenotemark[1]  & 31458 & 31406 &       & 80838 & 80686 &       & 32925 & 32849 &       & 82035 & 81842 \\
  \end{tabular}
\end{ruledtabular}{\scriptsize \tablenotemark[1] Ionization energy of the ground-state valence electron}%
\label{tab:ExpEnComp}%
\end{table*}%

\begin{table*}[ht!]
    \centering
    \caption{Calculated energy levels (BO) for the francium-like actinide ions and available experimental data~\cite{BWParis}. (cm$^{-1}$)}
    \label{tab:ActinideEnergy}
\begin{ruledtabular}
\begin{tabular}{lrr|lrr|lr|lrr|lr}
\multicolumn{3}{c|}{Ac${}^{2+}$}  & \multicolumn{3}{c|}{Th${}^{3+}$}  & \multicolumn{2}{c|}{Pa${}^{4+}$}  & \multicolumn{3}{c|}{U${}^{5+}$}      &   \multicolumn{2}{c}{Np${}^{6+}$}   \\
Level &  \multicolumn{1}{c}{BO}  &  \multicolumn{1}{c|}{Exp.}  &   & \multicolumn{1}{c}{BO}  &  \multicolumn{1}{c|}{Exp.}  &     &  \multicolumn{1}{c|}{BO}  &   &  \multicolumn{1}{c}{BO}  &  \multicolumn{1}{c|}{Exp.}  &   & \multicolumn{1}{c}{ BO} \\
\hline
   7s$_{1/2}$  & 0     & 0     & 5f$_{5/2}$  & 0     & 0     & 5f$_{5/2}$  & 0     &  5f$_{5/2}$  & 0     & 0     &  5f$_{5/2}$  & 0 \\
   6d$_{3/2}$  & 435   & 801   & 5f$_{7/2}$  & 4393  & 4325  & 5f$_{7/2}$  & 6061  &  5f$_{7/2}$  & 7784  & 7609  &  5f$_{7/2}$  & 9470 \\
   6d$_{5/2}$  & 3926  & 4204  & 6d$_{3/2}$  & 8681  & 9193  & 6d$_{3/2}$  & 48302 &  6d$_{3/2}$  & 90713 & 91000 &  6d$_{3/2}$  & 139415 \\
   5f$_{5/2}$  & 23467 & 23455 & 6d$_{5/2}$  & 14084 & 14486 & 6d$_{5/2}$  & 55753 &  6d$_{5/2}$  & 100369 & 100511 &  6d$_{5/2}$  & 151486 \\
   5f$_{7/2}$  & 26112 & 26080 & 7s$_{1/2}$  & 22948 & 23131 & 7s$_{1/2}$  & 79208 &  7s$_{1/2}$  & 141157 & 141448 &  7s$_{1/2}$  & 211402 \\
   7p$_{1/2}$  & 29375 & 29466 & 7p$_{1/2}$  & 59957 & 60239 & 7p$_{1/2}$  & 123396 &  7p$_{1/2}$  & 193744 & 193340 &  7p$_{1/2}$  & 273437 \\
   7p$_{3/2}$  & 38136 & 38063 & 7p$_{3/2}$  & 72995 & 73056 & 7p$_{3/2}$  & 141201 &  7p$_{3/2}$  & 216937 & 215886 &  7p$_{3/2}$  & 301680 \\
   8s$_{1/2}$  & 69660 &       & 8s$_{1/2}$  & 120357 & 119622 & 7d$_{3/2}$  & 201271 &  6f$_{5/2}$  & 283289 &       &  6f$_{5/2}$  & 365542 \\
   7d$_{3/2}$  & 73543 &       & 7d$_{3/2}$  & 120907 & 119685 & 7d$_{5/2}$  & 203789 &  6f$_{7/2}$  & 284244 &       &  6f$_{7/2}$  & 366832 \\
   7d$_{5/2}$  & 74579 &       & 7d$_{5/2}$  & 122622 & 121427 & 6f$_{5/2}$  & 203997 &  7d$_{3/2}$  & 288691 &       &  7d$_{3/2}$  & 382067 \\
   8p$_{1/2}$  & 80612 &       & 8p$_{1/2}$  & 135196 & 134517 & 6f$_{7/2}$  & 204665 &  7d$_{5/2}$  & 292124 &       &  7d$_{5/2}$  & 386532 \\
   6f$_{5/2}$  & 83166 &       & 8p$_{3/2}$  & 140536 & 139871 & 8s$_{1/2}$  & 205653 &  8s$_{1/2}$  & 299566 &       &  8s$_{1/2}$  & 400640 \\
   6f$_{7/2}$  & 83513 &       & 9s$_{1/2}$  & 161461 & 160728 & 8p$_{1/2}$  & 224369 &  8p$_{1/2}$  & 322245 &       &  8p$_{1/2}$  & 428393 \\
  Lim.:
& 141221 &  140590\tablenotemark[1]  &       & 232015 &  231065\tablenotemark[1]  &   & 363394 &       & 509109 &  500000\tablenotemark[1]  &       & 667359 \\
\end{tabular}
 \end{ruledtabular}{\scriptsize \tablenotemark[1] Theoretical values: $140590(160)$, $231065(200)$, $500000(13000)$~\cite{NIST}}%
\end{table*}%

  \begin{table}
    \centering%
    \caption{Percentage difference between calculations and experiment for the most important energy intervals for PNC.}%
\begin{ruledtabular}%
  \begin{tabular}{rrrrr}
  {Energy Interval} & \multicolumn{1}{c}{Fr} & \multicolumn{1}{c}{Ra$^+$} & \multicolumn{1}{c}{Ac$^{2+}$} & \multicolumn{1}{c}{Th$^{3+}$} \\
\hline
  7$s_{1/2}-7p_{1/2}$ & -0.2\% & -0.3\% & -0.3\% & -0.3\% \\
  7$s_{1/2}-8p_{1/2}$ & 0.3\% & 0.5\% & ---   & 0.8\%
  \end{tabular}%
\end{ruledtabular}%
    \label{tab:intervals}%
  \end{table}%

  \begin{table}
    \centering%
    \caption{Calculated ionisation energies (cm$^{-1}$) including ladder diagrams for La$^{2+}$ and comparison with experiment.}%
\begin{ruledtabular}%
  \begin{tabular}{lrrrrr}
 Level &\multicolumn{1}{c}{BO}& \multicolumn{1}{c}{Ladder}& \multicolumn{1}{c}{Sum} &\multicolumn{1}{c}{Exp.~\cite{NIST}}   & \multicolumn{1}{c}{\% Diff.}\\
\hline
$6s_{1/2}$&141301&-204&141097&141084&0.01\% \\
$6p_{1/2}$&112930&-189&112741&112660&0.07\% \\
$6p_{3/2}$&109780&-174&109606&109564&0.04\% \\
$5d_{3/2}$&155565&-1005&154561&154675&-0.07\% \\
$5d_{5/2}$&153902&-1025&152877&153072&-0.13\% \\
  \end{tabular}%
\end{ruledtabular}%
    \label{tab:La2+Ladder}%
  \end{table}%

\begin{table}
  \centering%
  \caption{Calculated reduced matrix elements of transitions of interest to $s$-$d$ PNC calculations. RPA equations solved at the energy of each PNC transition. (a.u.)} %
\begin{ruledtabular}%
    \begin{tabular}{rlll}
    Transition &Ra$^{+}$ & Ac$^{2+}$ & Th$^{3+}$  \\
\hline
    $7s_{1/2}-7p_{3/2}$ 		&4.485& 3.775   & 3.326    \\
    $7s_{1/2}-8p_{3/2}$ 		&0.3729& 0.1755  & 0.0770  \\
    $6d_{3/2}-7p_{1/2}$ 		&3.533& 2.569   & 2.100    \\
    $6d_{3/2}-8p_{1/2}$ 		&0.0440& 0.2047 & 0.2275%
    \end{tabular}%
  \label{tab:MEs}%
 \end{ruledtabular}%
\end{table}%

  \begin{table*}%
    \centering%
    \caption{Calculations of reduced E1 and E2 matrix elements   $\langle \psi_f||\tilde d_{Ek}||\psi_i\rangle$
 for the Fr-like actinide ions, including Brueckner correlations and core polarization
 (a.u.).\\ Note that the reduced matrix elements observe the symmetry property {$\langle  a||\hat d||b\rangle = (-1)^{J_a-J_b}\langle b ||\hat d|| a \rangle$}.}%
\begin{ruledtabular}
\begin{tabular}{ll|ddddddd}
   \multicolumn{2}{c|}{E1}  &  & &  & &  && \\
 \multicolumn{1}{c}{$i$} &  
\multicolumn{1}{c|}{$f$}  & 
\multicolumn{1}{c}{Fr} & 
\multicolumn{1}{c}{Ra$^+$} & 
\multicolumn{1}{c}{Ac$^{2+}$} & 
\multicolumn{1}{c}{Th$^{3+}$} & 
\multicolumn{1}{c}{Pa$^{4+}$} & 
\multicolumn{1}{c}{U$^{5+}$}  & 
\multicolumn{1}{c}{Np$^{6+}$}\\
\hline
  $7s_{1/2}$ & $7p_{1/2}$ & 4.287 & 3.228 & 2.707 & 2.377 & 2.147 & 1.969 & 1.830 \\
        & $7p_{3/2}$ & 5.906 & 4.482 & 3.771 & 3.316 & 2.995 & 2.745 & 2.546 \\
        & $8p_{1/2}$ & 0.287 & -0.071 & -0.196 & -0.261 & -0.306 & -0.342 & -0.370 \\
        & $8p_{3/2}$ & 0.890 & 0.360 & 0.148 & 0.026 & -0.060 & -0.131 & \multicolumn{1}{c}{---} \\
  $8s_{1/2}$ & $7p_{1/2}$ & -4.195 & -2.497 & -1.887 & -1.554 & -1.340 & -1.188 & -1.101 \\
        & $7p_{3/2}$ & -7.425 & -4.621 & -3.563 & -2.974 & -2.595 & -2.329 & -2.162 \\
        & $8p_{1/2}$ & 10.150 & 7.006 & 5.599 & 4.758 & 4.186 & 3.762 & 3.426 \\
        & $8p_{3/2}$ & 13.442 & 9.386 & 7.547 & 6.437 & 5.676 & 5.107 & 4.643 \\
  $6d_{3/2}$ & $7p_{1/2}$ & 7.174 & 3.533 & 2.571 & 2.096 & 1.803 & 1.587 & 1.433 \\
        & $7p_{3/2}$ & -3.301 & -1.496 & -1.050 & -0.837 & -0.708 & -0.614 & -0.549 \\
        & $8p_{1/2}$ & -2.489 & 0.041 & 0.189 & 0.185 & 0.13  & \multicolumn{1}{c}{---}   & \multicolumn{1}{c}{---} \\
        & $8p_{3/2}$ & 0.764 & -0.137 & -0.156 & -0.138 & \multicolumn{1}{c}{---}   & \multicolumn{1}{c}{---}   & \multicolumn{1}{c}{---} \\
  $6d_{5/2}$ & $5f_{7/2}$ & -9.117 & -5.666 & -2.819 & -1.930 & -1.498 & -1.217 & -0.996 \\
        & $7p_{3/2}$ & 10.156 & 4.795 & 3.417 & 2.749 & 2.344 & 2.044 & 1.835 \\
        & $8p_{3/2}$ & -2.499 & 0.379 & 0.487 & 0.453 & 0.366 & \multicolumn{1}{c}{---}  & \multicolumn{1}{c}{---} \\
  $5f_{5/2}$ & $6d_{3/2}$ & 7.318 & 4.441 & 2.173 & 1.510 & 1.184 & 0.968 & 0.795 \\
        & $6d_{5/2}$ & -2.037 & -1.257 & -0.604 & -0.417 & -0.327 & -0.271 & -0.229 \\
\hline\hline
   \multicolumn{2}{c|}{E2}  &  & &  & &  && \\
 \multicolumn{1}{c}{$i$} &  
\multicolumn{1}{c|}{$f$}  & 
\multicolumn{1}{c}{Fr} & 
\multicolumn{1}{c}{Ra$^+$} & 
\multicolumn{1}{c}{Ac$^{2+}$} & 
\multicolumn{1}{c}{Th$^{3+}$} & 
\multicolumn{1}{c}{Pa$^{4+}$} & 
\multicolumn{1}{c}{U$^{5+}$}  & 
\multicolumn{1}{c}{Np$^{6+}$}\\
\hline
  $7s_{1/2}$ & $6d_{3/2}$ & -33.367 & -14.676 & -9.529 & -7.065 & -5.612 & -4.591 & \multicolumn{1}{c}{---} \\
        & $6d_{5/2}$ & -41.568 & -18.868 & -12.373 & -9.225 & -7.359 & -6.049 & -3.9 \\
  $5f_{5/2}$ & $7p_{1/2}$ & 69.761 & 33.569 & 6.711 & 3.086 & 1.97  & 1.35  &\multicolumn{1}{c}{---} \\
        & $7p_{3/2}$ & -47.191 & -20.289 & -3.408 & -1.438 & \multicolumn{1}{c}{---}   & \multicolumn{1}{c}{---}   & \multicolumn{1}{c}{---} \\
  \end{tabular}%
\end{ruledtabular}%
    \label{tab:FrMEs}%
  \end{table*}%

The PNC  amplitude 
of a transition between states $a$ and $b$ of the same parity can be
expressed via the sum over opposite parity states $n$,  
\small\begin{equation}%
E_{PNC} = \sum_n 
	 \Big[ \frac{\langle
  b|\hat{d}_{E1}|{n}\rangle\langle{n}|\hat{h}_{W}|{a}\rangle}{E_{a}-E_{n}}   
	+\frac{\langle b|\hat{h}_{W}|n\rangle\langle
          n|\hat{d}_{E1}|{a}\rangle}{E_{b}-E_{n}}\Big], 
	\label{eq:pnc}
\end{equation}\normalsize
where $a$, $b$, and $n$ are many-electron wave functions of the atom,
$\hat{d}_{E1}$ is the electric dipole transition operator and
$\hat{h}_{W}=\tfrac{G_F}{2\sqrt{2}}Q_W\rho(r)\gamma_5$  
is the nuclear-spin--independent PNC interaction ($G_F$ the Fermi
constant,  $Q_W$ is the nuclear weak charge, $\rho$ the nucleon
density and the Dirac matrix $\gamma_5=i\gamma_0\gamma_1\gamma_2\gamma_3$).    

The exact expression (\ref{eq:pnc}) can be reduced to an approximate
one which looks very similar but contains single-electron matrix
elements and energies. This serves as a base for the {\em
  sum-over-states} calculations. Many-body effects are included in
this approach by modifying single-electron orbitals and the external
field operators. We use this approach for tests only. The actual
calculations are done using a different approach based on solving
differential equations. This approach has many important advantages
which will be discussed below. 

We start from the mean-field Dirac-Fock approximation with a $V^{N-1}$
potential, which is particularly convenient for the single valence
electron ions studied here, and then include dominating electron
correlation effects. Core-valence correlation corrections to the
valence electron  wavefunction are included via the correlation
potential method~\cite{CPM}. The correlation potential, $\hat \Sigma$,
including summation of the series of dominating diagrams, is
calculated in all orders of perturbation theory using relativistic
Green functions and the Feynman diagram technique~\cite{CPM2}.  
The correlation potential $\hat \Sigma$  is then used to construct the
so-called {\em Brueckner orbitals} (BOs) for the valence electron. 
The BOs are found by solving the relativistic Hartree-Fock-like
equations with the extra operator $\hat \Sigma$:   
 $(\hat H_0 +\hat \Sigma - E_n)\psi_n^{(\rm BO)}=0$,
where $\hat H_0$ is the relativistic Hartree-Fock Hamiltonian and the
index $n$ denotes valence states. The BO $\psi_n^{(\rm BO)}$ and
energy $E_n$ include correlations.    

The weak and electric dipole interactions and the electron core
polarization effects are included via the time-dependent Hartree-Fock
approximation~\cite{CPM,CPM2}, which is equivalent to the well-known
{random phase approximation} (RPA). 
The single-electron wavefunction in an external weak and $E1$ field is
expressed in the RPA method as 
$\psi = \psi_0 + \delta\psi +X e^{-i\omega t}+Y e^{i\omega t} + \delta
Xe^{-i\omega t} + \delta Ye^{i\omega t}$,
where $\psi_0$ is the unperturbed state, $\delta\psi$ is the
correction due to weak interaction acting alone, $X$ and $Y$ are
corrections due to the photon field acting alone, $\delta X$ and
$\delta Y$ are corrections due to both fields acting simultaneously,
and $\omega$ is the frequency of the PNC transition. Where possible,
$\omega$ should be taken from the experimental energy of the
transition. The corrections are found by solving the system of RPA
equations self-consistently for the core: 
\begin{align}
(\hat H_0 - E_c)\delta\psi_c &= - (\hat h_W + \delta \hat
V_W)\psi_{0c}, \nonumber \\ 
(\hat H_0 - E_c-\omega)X_c &= - (\hat d_{E1} + \delta \hat
V_{E1})\psi_{0c}, \nonumber \\ 
(\hat H_0 - E_c+\omega)Y_c &= - (\hat d_{E1}^{\dagger} + \delta \hat
V_{E1}^{\dagger})\psi_{0c}, \label{eq:RPA}\\
(\hat H_0 - E_c-\omega)\delta X_c &= -\delta\hat V_{E1}\delta\psi_c -
\delta\hat V_WX_c - \delta\hat V_{E1W}\psi_{0c}, \nonumber \\ 
(\hat H_0 - E_c+\omega)\delta Y_c &= -\delta\hat
V_{E1}^{\dagger}\delta\psi_c -
\delta\hat V_WY_c - \delta\hat V_{E1W}^{\dagger}\psi_{0c}, \nonumber 
\end{align}
where the index $c$ denotes core states, $\delta \hat V_W$ and $\delta
\hat V_{E1}$ are corrections to the core potential arising from the
weak and E1 interactions respectively and $\delta \hat V_{E1W}$ is the
correction to the core potential arising from the simultaneous
perturbation of the weak field and the electric field of the laser
light. 

The PNC amplitude between valence states $a$ and $b$ in the RPA
approximation is given by 
\begin{align}
 E_{\rm PNC} 
&= \langle \psi_b|\hat d_{E1} +  \delta\hat V_{E1}|\delta\psi_a\rangle 
+ \langle \psi_b|\hat h_{W} + \delta\hat V_{W}|X_a\rangle \notag  \\ 
&+ \langle \psi_b|\delta\hat V_{E1W}|\psi_a\rangle \label{amp}  \\
&= \langle \psi_b|\hat d_{E1} + \delta\hat V_{E1}|\delta\psi_a\rangle
+  \langle \delta\psi_b|\hat d_{E1} + \delta\hat V_{E1}|\psi_a\rangle \notag \\ 
&+ \langle \psi_b|\delta\hat V_{E1W}|\psi_a\rangle.  \notag
\end{align}
By using BOs for the valence states $\psi_a$ and $\psi_b$ in
(\ref{amp}) we can include correlations in the calculation of the PNC
amplitude. The corrections $\delta \psi_a$ and $\delta \psi_b$ to the
BOs $\psi_a$ and $\psi_b$ are also found with the use of the
correlation potential $\hat \Sigma$: 
\begin{equation}
(\hat H_0 -E_a + \hat \Sigma)\delta\psi_a = -(\hat h_W + \delta\hat
V)\psi_{0a}. \label{eq:dpsiv} 
\end{equation}

Note that the last term in (\ref{amp}) gives an important contribution
that is usually not included in sum-over-states calculations. It
represents the {\em double core-polarization} by simultaneous action
of two external fields; the electric field of laser light and weak
electron-nucleus interaction. 

This term is negligible for the $6s$-$7s$ PNC transition in Cs by
chance only, and is very different for other atoms and transitions.  
It is 2\% for the $6s$-$5d$ PNC transition of Cs, 5\% for the
$7s$-$6d$ transition in Ra$^+$ and 40\% for the $6p_{1/2}$-$6p_{3/2}$
transition in Tl. The last applies to the case when thallium is
treated as a one-valence-electron system, so that the $6s$ electrons
remain in the core. The reason why it is not usually included in
sum-over-states calculations is that it cannot be represented as a
product of single-electron matrix elements involving valence states.  
The problem of double core polarization will be considered in more
detail elsewhere~\cite{double}.


Table~\ref{tab:ExpEnComp} presents our calculated energy levels for
Cs, Ba$^{+}$, Fr and Ra$^+$  against experimental values. The BO
calculations are accurate to around $0.1$-$0.5\%$ for most levels,
which is typical for this type of calculation.  
Table~\ref{tab:ActinideEnergy} presents our calculated energy levels
for the francium-like actinide ions,  
 and Table~\ref{tab:intervals} shows the percentage difference between
 our calculations and experimental values for the most important
 energy intervals for PNC in Fr, Ra$^{+}$, Ac$^{2+}$ and Th$^{3+}$.  

Note that the accuracy can be further improved by including the
contributions of the so-called {\em ladder diagrams}~\cite{ladder}. We
illustrate thsi using the La$^{2+}$ ion as an example.
Table~\ref{tab:La2+Ladder} presents calculations of La$^{2+}$
ionisation energies including ladder diagrams. The experimental
energies are reproduced to an extraordinary accuracy, even for the
notoriously difficult $d$ levels. A full inclusion of the ladder
diagrams for all ions will be saved for a later work, and currently
our method only allows inclusion of ladder diagrams in the energy
levels but not PNC. Here we demonstrate that by including the ladder
diagrams the accuracy is significantly improved, and that the accuracy
in all of these ions is very good.

We also calculate  several reduced E1 matrix elements that are of
interest to PNC transitions, which are presented in
Table~\ref{tab:MEs}, and in Table~\ref{tab:FrMEs} we present
calculations of several of the reduced matrix elements of the
considered Fr-like actinide ions.


In francium, the $7s$ state is the ground state. However, in charged
ions this is not necessarily the case. For the ions after actinium the
$5f$ state is pushed below $7s$, forming a new ground state (see
Table~\ref{tab:ActinideEnergy}). Also, after actinium the $6d_{3/2}$
state is pushed below the $7s$ state. The ions after Np$^{6+}$ no
longer have closed $p$-shells and are not considered here.  
A similar crossing of configurations also occurs in the cesium
isoelectronic sequence; Cs and Ba$^+$ have 6$s$ ground-states,
La$^{2+}$ has 5$d_{3/2}$, and Ce$^{3+}$ and Pr$^{4+}$ have 4$f$
ground-states (see Tables~\ref{tab:ExpEnComp}
and~\ref{tab:La2+Ladder}). 

For a $7s$-$6d_{3/2}$ interval to be a viable transition for the
measurement of PNC, one of these states ($7s$ or $6d_{3/2}$) should be
either the ground state or a metastable state that can first be
populated and then the PNC transition observed.

Also, it was shown it the pivotal work of N Fortson~\cite{BaII} that
to ensure accurate PNC measurements of a single trapped ion both the
upper and lower levels of the transition should be long lived.  In
Table~\ref{tab:lifetimes} we present calculations of the lifetimes of
the relevent levels for Ba$^{+}$, La$^{2+}$, Ra$^{+}$, Ac$^{2+}$ and
Th$^{3+}$.  
We show that this condition is met in all of these ions except for
Th$^{3+}$, which has a long-lived upper level but a lower level that
quickly decays via E1 transitions. 

Note in particular the very long-lived upper ($6d_{3/2}$) state of
Ac$^{2+}$. This state is practically stable, the E2 transition back to
the $7s$ ground state (the only lower state - see
Table~\ref{tab:ActinideEnergy}) is highly suppressed due to the very
small energy gap of this state, $801$ cm$^{-1}$. 
This is very benefitial for the measurent of PNC in single-trapped
ions~\cite{BaII}.

  \begin{table}%
    \centering%
    \caption{Lifetimes (s) of upper and lower states of the $s$-$d$
      PNC transitions for main ions of interest, where $n$ is the
      principal quantum number: $n=6$ for Ba$^{+}$ and La$^{2+}$,
      $n=7$ for Ra$^{+}$, Ac$^{2+}$ and Th$^{3+}$. ($\infty$ means
      ground state).}%
\begin{ruledtabular}%
\begin{tabular}{llllll}%
	Level&\multicolumn{1}{c}{Ba$^{+}$}	&\multicolumn{1}{c}{La$^{2+}$}&\multicolumn{1}{c}{	Ra$^{+}$}&	\multicolumn{1}{c}{Ac$^{2+}$}	&\multicolumn{1}{c}{Th$^{3+}$}\\
\hline
$(n$-$1)d_{3/2}$&	84.5	&$\infty$&	0.642	&1.19$\times10^{6}$&	0.58\\
$ns_{1/2}$&	$\infty$	&0.347&	$\infty$&$\infty$&	1.12$\times10^{-6}$
\end{tabular}%
\end{ruledtabular}%
    \label{tab:lifetimes}%
  \end{table}%

\section{Results and discussion}

The final calculations of the $s$-$d$ and (near) optical $s$-$s$ PNC
amplitudes for the francium-like ions are presented in
Table~\ref{tab:actinides} with some previous calculations for
comparison. The amplitudes calculated here include core-polarization
(RPA) and all-order Brueckner correlations.  

For comparison and completeness, these calculations were also
performed for Cs, Ba$^+$ and the first few Cs-like lanthanide ions.
These much lighter ions have correspondingly smaller PNC
amplitudes. The results are presented in Table~\ref{tab:lanthanides}.   
We have not presented a result for the $6s$-$7s$ transition in cesium
since this has been investigated thoroughly in our recent
work~\cite{CsPNC}.

The PNC amplitudes calculated here agree very well with previous
determinations for Cs, Ba$^+$, Fr and Ra$^+$. 
For Ra$^+$ our result is within 1\% of the result calculated in
Ref.~\cite{Ba+} using the same `solving equations' method used
here~\cite{isotopes}.  
Our Ra$^+$ value is also, however, 4-5\% smaller than the amplitudes
calculated in that same work as well as in Ref.~\cite{SafRa+}, which
used a different `sum-over-states' approach. The difference is most
likely due to the {\em double core-polarization} (last term of
(\ref{amp}), discussed above), which we calculate to contribute
$-4.7\%$ to this amplitude, and is not included in the sum-over-states
calculations. Note that {double core-polarization} was also not
included in our recent calculations for Ba$^+$, Yb$^+$, and
Ra$^+$~\cite{Yb+}.  This is because in that paper we were focused on
the nuclear spin-dependent PNC amplitudes, where high accuracy of the
analysis is less important.

  \begin{table}[t]%
    \centering%
    \caption{Calculated  $7s$-$6d_{3/2}$ 
and  $7s$-$8s$ 
PNC amplitudes 
for the Fr-like actinide ions, in units $i(-Q_W/N)\times 10^{-11}$
a.u. Also shown are the ground-state levels, experimental wavelengths
of the transitions, and several previous PNC calculations for
comparison.}%
\begin{ruledtabular}%
\begin{tabular}{llrlddr}%
Ion   &  \multicolumn{1}{c}{ground} &   \multicolumn{2}{c}{$\lambda$ (nm)} &  \multicolumn{3}{c}{$E_{PNC}$}  \\
      &    \multicolumn{1}{c}{-state}    &       &       &  \multicolumn{1}{r}{This work} &  \multicolumn{2}{c}{Others} \\
\hline%
      ${}^{223}$Fr 			 & $7s_{1/2}$ & $sd$  & 616   & 57.99 & 57.1(23)&\cite{Ba+} \\
      &       & $ss$  & 507   &    15.38    &  15.56(17)\tablenotemark[1]&\cite{FrSaf} \\
      &       &   &    &  &  15.69(16)\tablenotemark[2]&\cite{ShabPRA} \\
      ${}^{226}$Ra${}^{+}$	 & $7s_{1/2}$ & $sd$  & 827   & 44.35 & 43.9(17)\tablenotemark[3]&\cite{Ba+} \\
      &       &       &       &       & 46.4& \cite{KVI} \\
      &       &       &       &       & 46.47\tablenotemark[1]& \cite{SafRa+} \\
      &       & $ss$  & 230   &  10.89 &&  \\
      ${}^{227}$Ac${}^{2+}$	 & $7s_{1/2}$ & $sd$  & 12484 & 42.81 &&  \\
      ${}^{232}$Th${}^{3+}$	 & $5f_{5/2}$ & $sd$  & 717   & 43.59 & & \\ 
      ${}^{231}$Pa${}^{4+}$	 & $5f_{5/2}$ & $sd$  & $324$\tablenotemark[4] &  43.49 &&  \\
      ${}^{238}$U${}^{5+}$	 & $5f_{5/2}$ & $sd$  & 198   &  45.94 &  &\\
      ${}^{237}$Np${}^{6+}$	 & $5f_{5/2}$ & $sd$  &  $139$\tablenotemark[4]  &  44.07 &&%
\end{tabular}%
\tablenotetext[1]{Contribution of Breit interaction is removed for the convenience of comparison. The final result of \cite{FrSaf} is -15.41, and of \cite{SafRa+} is 45.89}
\tablenotetext[2]{Breit and QED corrections removed for comparison. The final result of \cite{ShabPRA} is 15.49}
\tablenotetext[3]{Rescaled from $^{223}$Ra \cite{isotopes}}
\tablenotetext[4]{Calculated (BO) wavelength}
\end{ruledtabular}%
    \label{tab:actinides}%
  \end{table}%

  \begin{table}%
    \centering%
    \caption{Calculations of the PNC amplitudes for the Cs-like actinide ions and several previous PNC calculations for comparison. In units $i(-Q_W/N)\times 10^{-11}$ a.u.}%
\begin{ruledtabular}%
\begin{tabular}{llrlddr}%
Ion   &  \multicolumn{1}{c}{ground} &   \multicolumn{2}{c}{$\lambda$ (nm)} &  \multicolumn{3}{c}{$E_{PNC}$}  \\
      &    \multicolumn{1}{c}{-state}    &       &       &  \multicolumn{1}{c}{This work} &  \multicolumn{2}{c}{Others} \\
\hline%
      ${}^{133}$Cs     &    $6s_{1/2}$	     & $sd$  & 690   & 3.703 & 3.62(14)&\cite{Ba+} \\
         ${}^{137}$Ba${}^{+}$&$6s_{1/2}$          & $sd$  & 2051  & 2.197 & 2.17(9)&\cite{Ba+} \\ 
      &       &       &       &       &2.46(2)&\cite{SahBa+} \\
      &      & $ss$  & 236   &  0.6582 &  &\\ 
      ${}^{139}$La${}^{2+}$     &    $5d_{3/2}$     & $sd$  & 736   & 2.135 & & \\
      ${}^{140}$Ce${}^{3+}$     &     $4f_{5/2}$    & $sd$  & 721   & 2.076 &&  \\
      ${}^{141}$Pr${}^{4+}$     &      $4f_{5/2}$   & $sd$  & 156   & 2.102 &&%
\end{tabular}%
\end{ruledtabular}%
    \label{tab:lanthanides}%
  \end{table}%

The $sd$ PNC transitions tend to have a single dominating term which
contributes $\sim90\%$ to the total amplitude~\cite{Ba+}. In
Th$^{3+}$, for example, this term with the $7s$-$7p_{1/2}$ energy
interval contributes approximately $96\%$. 
The energy interval for this term agrees with experiment to 0.3\% (see
Table \ref{tab:intervals}).  
Based on comparison with experimental energies and previous
calculations, we expect our amplitudes to be accurate to around 1\%.  
This accuracy can be improved by including the Breit~\cite{Breit},
neutron-skin~\cite{skin} and QED~\cite{QED} corrections, missed
high-order correlations such as ladder
diagrams~\cite{ladder} (see Table~\ref{tab:La2+Ladder}) and structural
radiation~\cite{CPM2}, and  with the use of experimental $p$-$d$ E1
amplitudes.  
With these corrections the theoretical accuracy can be expected to
surpass that of cesium. 

The experimental accuracy can be expected to be high due to stable
nuclei and large PNC signals.  
Additionally, in the case of Ac$^{2+}$ where both upper and lower
levels are extremely long-lived, the experimental accuracy has the
potential to be very high. 

\section{Accessing the PNC transition.}

\begin{figure}[t]
\centering
\caption{Level scheme for Th$^{3+}$}
\includegraphics[width=0.33\textwidth]{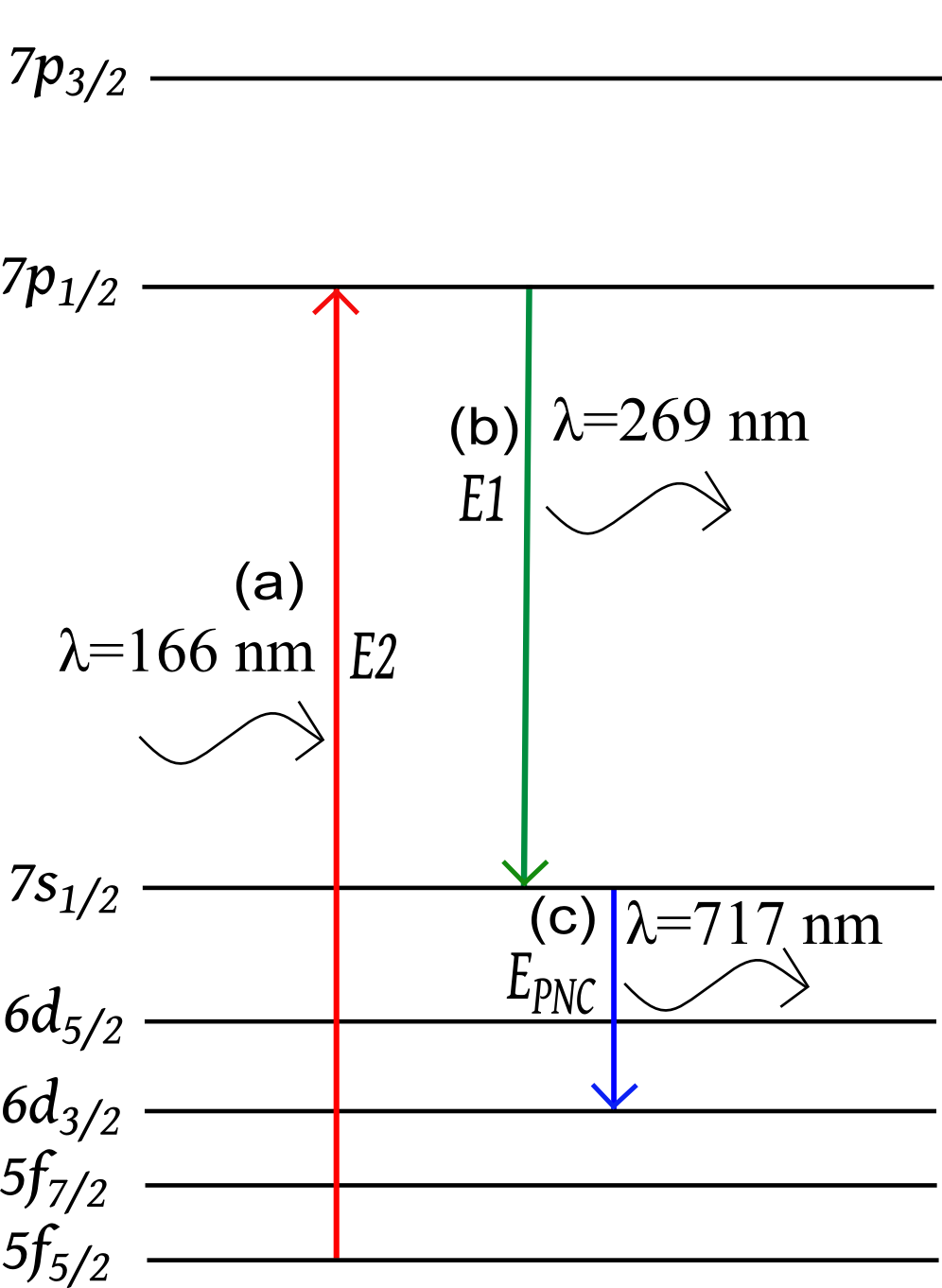}
\label{fig:ThLevel}%
\end{figure}

In order to observe the $7s$-$6d_{3/2}$ PNC transitions in the
actinide ions with the $5f_{5/2}$ ground states, the $7s$ state must first
be populated. In these ions the $5d_{3/2}$ state lies below the $7s$
state, however it is unstable as it will decay very quickly via an E1
transition to the $5f_{5/2}$ ground state.  Population of the $7s$
state can be achieved via optical excitation to the 7$p_{1/2}$ or
7$p_{3/2}$ levels by a series of E1 transitions (e.g. $5f$-$6d$-$7p$)
or an E2 transition,  then 7$p_{1/2,3/2}$ will spontaneously decay to
the 7$s$ state via an E1 transition -- see Fig.~\ref{fig:ThLevel}.

For this method to be viable we need to meet several criteria. First,
we need the 7$s$ state to be metastable and have an appropriate
lifetime. Second, we should also have that the pumping transition
frequencies (to populate $7s$) are in the range of laser spectroscopy.   
Also, it is necessary that the de-excitation from the $p$ to $s$-level
is relatively probable compared to transitions to other levels. If
this last condition is not met it is possible to enforce it using
stimulated emission, which should not be a problem since these
transitions are optical.  

Table~\ref{tab:prob} shows the probabilities and per-second transition
rates for these transitions in Th$^{3+}$. Our calculations are in very
good agreement with existing calculations of Safronova {\em et
  al.}~\cite{SafPRA06,SafArXiv}.  
The 7$s$ state should be relatively stable, since there are no allowed
E1 transitions to lower states. The only significant contributions to
its decay are from E2 transitions to the $6d_{3/2}$ and $6d_{5/2}$
states. We calculate a long lifetime of 0.58 s, in excellent agreement
with the recent calculation of M. S. and U. I. Safronova of 0.570(8)
s~\cite{SafArXiv}.

\begin{table}[h!]%
  \centering
  \caption{Energies ($\omega$), probabilities ($\Gamma$)  and per-second transition rates ($A_r$) of transitions in Th$^{3+}$.} 
\begin{ruledtabular}
\begin{tabular}{lllll}
\multicolumn{2}{c}{Transition}       & \multicolumn{1}{c}{$\omega$ (a.u.)}  & \multicolumn{1}{c}{$\Gamma$  (a.u.)} &\multicolumn{1}{c}{$A_r$}(s${}^{-1}$) \\
\hline
$5f_{5/2}-6d_{3/2}$ & E1    	& 0.042 	& 1.46$\times10^{-11}$ & 6.07$\times10^{5}$  \\ 
$6d_{3/2}-7p_{1/2}$ & E1    	& 0.233	& 7.21$\times10^{-9}$ & 2.99$\times10^{8}$ \\ 
\hline
$5f_{5/2}-7p_{1/2}$ & E2    	& 0.274 	& 3.41$\times10^{-15}$ & 142 \\ 
$7p_{1/2}-7s_{1/2}$ & E1    		& 0.169  	& 7.07$\times10^{-9}$ &2.92$\times10^{8}$ \\  
$7p_{1/2}-6d_{3/2}$ & E1    	& 0.233  	& 1.45$\times10^{-8}$ &5.97$\times10^{8}$\\ 
\hline
$5f_{5/2}-7p_{3/2}$ & E2    	& 0.333  	& 1.96$\times10^{-15}$ &81.4\\
$7p_{3/2}-7s_{1/2}$ & E1    		& 0.227 	& 1.66$\times10^{-8}$ &6.90$\times10^{8}$\\  
$7p_{3/2}-6d_{3/2}$ & E1    	& 0.291 	& 2.24$\times10^{-9}$ &9.28$\times10^{7}$\\ 
$7p_{3/2}-6d_{5/2}$ & E1    	& 0.267 	& 1.87$\times10^{-8}$ &7.30$\times10^{8}$\\ 
\hline
$7s_{1/2}-6d_{3/2}$ & E2    	& 0.064 	&  3.57$\times10^{-17}$  & 1.48 \\  
$7s_{1/2}-6d_{5/2}$ & E2    	& 0.039 	&  5.57$\times10^{-18}$  & 0.23%
\end{tabular}\end{ruledtabular}\label{tab:prob}\end{table}%

\section{Conclusion}

We have provided calculations of parity nonconservation, energy
levels, matrix elements and lifetimes of several Cs-like rare-earth
and Fr-like actinide ions. 

We demonstrate that these systems provide a very high theoretical
accuracy. With the inclusion of other effects (such as Breit, QED,
ladder operators etc.) this could lead to better precision in the
calculations than has been achieved in cesium. 

With very large PNC amplitudes, these ions can be expected also to
have a very high accuracy in the measurements, with the added benefits
of (near)-stable nuclei in Ba$^{+}$, La$^{2+}$, Ac$^{2+}$ and
Th$^{2+}$, and long-lived upper and lower states for Ba$^{+}$,
La$^{2+}$, Ra$^{+}$ and particularly Ac$^{2+}$.

\acknowledgments
This work was supported by the Australian Research Council.



\begin{thebibliography}{40}
\bibitem{isotopes} The information about isotope numbers is missed in our old paper~\cite{Ba+}. It is most likely $^{223}$Ra in the mixed states calculations and $^{226}$Ra in the sum-over-states calculations.
\end{thebibliography}
\end{document}